%% file: root.tex
\documentclass[conference,a4paper, 9pt]{./IEEEtran}
\ifCLASSINFOpdf
\else
\fi
\usepackage{graphics} 
\usepackage[cmex10]{amsmath}
\usepackage{amsfonts, amssymb}
\usepackage{booktabs} 
\usepackage{ifthen}
\usepackage{tikz,pgfplots,pgfplotstable}
\graphicspath{{images/}}
\usepackage[nolist]{acronym}
\usepackage{siunitx}
\DeclareSIUnit\px{px}
\usepackage{cite}
\usepackage[hidelinks]{hyperref}
\providecommand{\myvec}[1]{\ensuremath{\boldsymbol{#1}}}

\providecommand{\undistDom}{\ensuremath{X}}
\providecommand{\distDom}{\ensuremath{Y}}
\providecommand{\distFct}{\ensuremath{\mathcal{C}}}
\providecommand{\distFctLearned}{\ensuremath{\tilde{\distFct}}}
\providecommand{\generatorForward}{\ensuremath{G}}
\providecommand{\generatorBackward}{\ensuremath{F}}
\providecommand{\discriminatorDist}{\ensuremath{D_\distDom}}
\providecommand{\discriminatorUndist}{\ensuremath{D_\undistDom}}
\newcommand{\discrOneLay}{\ac{ESD}}
\newcommand{\discrTwoLay}{\ac{SD}}
\newcommand{\discrThreeLay}{\ac{DAIS}}
\newcommand{\arruda}{\ac{CDOD}}
\providecommand{\blurQP}{\ensuremath{\sigma_\mathrm{blur}}}
\providecommand{\noiseQP}{\ensuremath{\sigma_\mathrm{AWGN}}}

\providecommand{\jpegQP}{\ensuremath{\mathrm{CL}}}

\usetikzlibrary{positioning,shadows,shapes,backgrounds}
\pgfplotsset{compat=1.17}
\usepgfplotslibrary{groupplots}
\pgfplotsset{every axis label/.append style={font=\small}}
\pgfplotsset{every tick label/.append style={font=\footnotesize}}
\pgfplotsset{every axis legend/.append style={font=\small}}
\pgfplotsset{every axis plot/.append style={line width=1.2pt}}
\pgfkeys{/pgf/number format/set thousands separator={\,}}
\pgfsetplotmarksize{1.5pt}
\tikzset{
	every picture/.style={>=latex},
	every node/.append style={font=\footnotesize},
	every pin/.append style={font=\tiny},
	every pin edge/.append style={shorten >=1pt, shorten <=.5pt},
}
\tikzstyle{syssplit} = [
circle,
fill=black,
draw=black,
inner sep=1pt,
]
\tikzstyle{syscon} = [%
draw,
shape=circle,
inner sep=0pt,
minimum width=4pt,
fill=white,
]
\tikzstyle{system}   = [draw, shape=rectangle, inner sep=3pt, align=center] 
\tikzstyle{source}   = [draw, shape=ellipse, inner sep=3pt] 
\definecolor{myred}{HTML}{ff0000}
\definecolor{myblue}{HTML}{0000ff}
\definecolor{mygreen}{HTML}{4daf4a}
\definecolor{mylila}{HTML}{984ea3}
\definecolor{myorange}{HTML}{ff7f00}
\definecolor{myyellow}{HTML}{ffff33}
\definecolor{mybrown}{HTML}{a65628}
\definecolor{mypink}{HTML}{f781bf}
\definecolor{mygray}{HTML}{999999}
\definecolor{mylightgray}{HTML}{eeeeee}
\definecolor{mygridcolor}{HTML}{d9d9d9}
\usepackage[firstpage=true]{background}
\SetBgContents{%
	\fbox{%
		\parbox{\textwidth}{%
			\footnotesize \textcopyright 2025 IEEE. Personal use of this material is permitted. Permission from IEEE must be obtained for all other uses, in any current or future media, including reprinting/republishing this material for advertising or promotional purposes, creating new collective works, for resale or redistribution to servers or lists, or reuse of any copyrighted component of this work in other works. DOI: \href{https://doi.org/10.1109/VCIP67698.2025.11396816}{10.1109/VCIP67698.2025.11396816} }}%
}	
\SetBgScale{1}
\SetBgAngle{0}
\SetBgPosition{current page.north}
\SetBgVshift{-1.1cm}
\SetBgColor{black}
\SetBgOpacity{1}
\begin{document}
\title{Domain Adaptation for Camera-Specific Image Characteristics using Shallow Discriminators\vspace*{-.6em}}
\author{\IEEEauthorblockN{Maximiliane Gruber, J\"{u}rgen Seiler, and Andr\'{e} Kaup}
	\IEEEauthorblockA{Multimedia Communications and Signal Processing\\
		Friedrich-Alexander-Universit\"{a}t Erlangen-N\"{u}rnberg (FAU), Erlangen, Germany\\
		Email: \{maximiliane.gruber, juergen.seiler, andre.kaup\}@fau.de}
}
\maketitle
%
\begin{acronym}
	\acro{AWGN}{additive white Gaussian noise}
	\acro{CL}{compression level}
	\acro{DA}{Domain Adaptation}
	\acro{DAIS}{domain adaptation in instance segmentation}
	\acro{CDOD}{cross-domain object detection}
	\acro{GAN}{Generative Adversarial Network}
	\acro{I2I}{image-to-image translation}
	\acro{HEIF}{High Efficiency Image File Format}
	\acro{JPEG}{Joint Photographic Experts Group}
	\acro{R-CNN}{Region-based Convolutional Neural Network}
	\acro{mAP}{mean Average Precision}
	\acro{PSNR}{Peak Signal-to-Noise Ratio}
	\acro{QP}{quantization parameter}
	\acro{BD-A}{basic discriminator}
	\acro{SD}{shallow discriminator}
	\acro{ESD}{extremely shallow discriminator}
\end{acronym}
\begin{abstract}
	Each image acquisition setup leads to its own camera-specific image characteristics degrading the image quality.
	In learning-based perception algorithms, characteristics occurring during the application phase, but absent in the training data, lead to a domain gap impeding the performance.
	Previously, pixel-level domain adaptation through unpaired learning of the pristine-to-distorted mapping function has been proposed.
	In this work, we propose shallow discriminator architectures to address limitations of these approaches.
	We show that a smaller receptive field size improves learning of unknown image distortions by more accurately reproducing local distortion characteristics at a low network complexity.
	In a domain adaptation setup for instance segmentation, we achieve mean average precision increases over previous methods of up to \num{0.15} for individual distortions and up to \num{0.16} for camera-specific image characteristics in a simplified camera model.
	In terms of number of parameters, our approach matches the complexity of one state of the art method while reducing complexity by a factor of \num{20} compared to another, demonstrating superior efficiency without compromising performance.
\end{abstract}
%
%
%
\acresetall
\vspace*{-.4em}
\section{Introduction}
\label{sec:introduction}
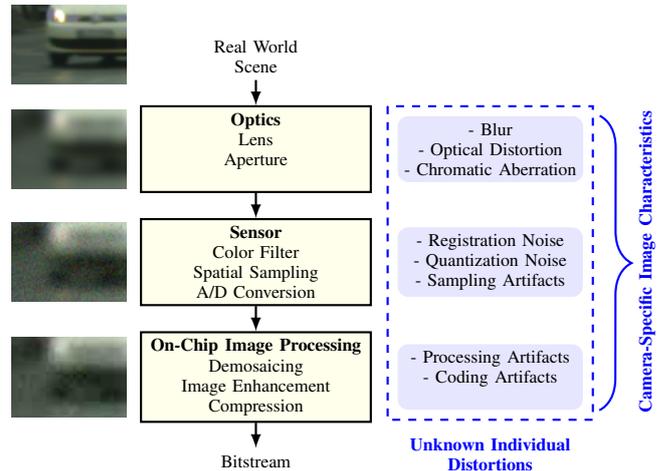
\begin{figure}
	\centering
	\input{images/unknown_distortion.tex}
	\vspace*{-1.4em}
	\caption{Camera-specific image characteristics result from the unknown, unique combination of individual image distortions (blue boxes) occurring along the imaging chain (yellow boxes).} 
	\vspace*{-1.7em}
	\label{fig:unknown-distortion}
\end{figure}

%
Perception algorithms in machine vision systems depend on digital cameras to capture their surroundings.
The hardware components and on-chip processing forming the image influence its appearance by multiple distortions, which are unknown in practice.
In \autoref{fig:unknown-distortion}, these distortions leading to the cameras unique combination of \textit{camera-specific image characteristics} are illustrated.
On-chip processing entails proprietary algorithms and a custom configuration of image coding tools and parameters. 
Thus, they contribute to camera-specific image characteristics through their implementation, although the underlying methods are not inherently camera-specific.
Camera-specific image characteristics cause a domain gap between datasets captured with different camera setups,
challenging the robust performance of learning-based perception.
The robustness against image distortions has been studied, 
e.g., for 
object detection~\cite{Geirhos2018, Fischer2021}, 
semantic segmentation~\cite{Kamann2021, Endo2023, Schiappa2024}, 
and
instance segmentation~\cite{Fischer2021, Schiappa2024}.
In~\cite{Schiappa2024}, continued relevance in visual foundation models is demonstrated.

Addressing the domain gap by obtaining annotated data for all target domains, i.e., every camera setup, is infeasible due to the associated cost and time.
Therefore, domain gaps are commonly addressed by \ac{DA}~\cite{Oza2024}.
In \ac{DA}, the mapping from source domain to target domain is learned from unlabeled data. 
Then, data from the annotated source domain is mapped to the target domain~\cite{Oza2024}.
Pixel-level \ac{DA} approaches the problem in the visual domain, often employing \ac{I2I}.
Paired \ac{I2I} requires source and target domain versions of the same images, for unpaired approaches disjoint sets of images suffice.
In contrast to image restoration techniques like super-resolution, pixel-level \ac{DA} has the advantage that it does not attempt to reconstruct lost information, but interprets the target domain data despite distortions.

Paired~\cite{Chen2020a, Ameur2023} and unpaired~\cite{Gruber2022} translation has been utilized to learn pristine-to-distorted mappings from an undistorted (pristine) source domain $\undistDom$ to a distorted target domain $\distDom$.
Individual distortions including blur, \ac{AWGN}, pink noise, JPEG, JPEG2000 and film grain are studied in~\cite{Chen2020a, Ameur2023, Gruber2022}.
In~\cite{Arruda2022}, the mapping between two camera systems is learned from unpaired data.
In addition,~\cite{Gruber2022} demonstrates improved \ac{DA} performance for instance segmentation and~\cite{Arruda2022} for object detection.
%

In~\cite{Gruber2022}, very good results are observed for the global distortions blur and \ac{AWGN}, while reproducing distortions caused by image coding shows limitations with their method \textit{\ac{DAIS}}.
For image coding, local distortion characteristics and content-dependent artifacts like banding, ringing and blockiness caused by adaptive block sizes are predominant.
Therefore, the limitations for image coding likely stem from the challenge of effectively capturing the local characteristics of image distortions and their content-dependency.
In~\cite{Arruda2022}, the author's pixel-level \ac{DA} approach \textit{\ac{CDOD}} has shown better results than state of the art feature-level \ac{DA} methods in a cross-camera setup.
During training, images are resized to $572 \times 572$ pixels, which alters their aspect ratio.
We suspect that learning camera-specific image characteristics is limited in this approach due to this non-uniform resizing operation.
Resizing changes the true distortions characteristics leading to undesired artifacts in the training images.

In this work, we build on \discrThreeLay{} addressing its limitations and introducing an improved method using shallow discriminators.
Our main contributions are:
We introduce a \ac{SD} and an \ac{ESD} with small receptive field sizes, hypothesizing that such architectures better capture local distortion characteristics at a low network complexity. 
We evaluate our methods by means of pixel-level \ac{DA} in instance segmentation and visual examples.
We further introduce and investigate a simplified camera model to extend our study to camera-specific image characteristics beyond unknown individual distortions.
We show the generalizability of our method in an ablation study.

\section{Proposed Method}
\label{sec:method}
We propose an unpaired pixel-level \ac{DA} approach for domain gaps due to camera-specific image characteristics.
Pixel-level approaches offer the advantage of being task-agnostic, i.e., adapting the dataset to the target domain is independent from the perception algorithm.
Additionally, the use of unpaired data ensures real-world applicability, as paired data is unattainable in practice.
Our method consists of three steps:
\vspace*{-.7em}
\begin{enumerate}
	\item Unpaired learning of unknown image distortion from test dataset
	\item Adaptation of pristine training dataset
	\item Adaptation of perception algorithm
\end{enumerate}
\vspace*{-.5em}

We aim to learn the true pristine-to-distorted mapping function $\distFct : \undistDom \rightarrow \distDom$ from source domain $\undistDom$ containing pristine data to target domain $\distDom$ containing distorted data.
We learn the pristine-to-distorted mapping function \distFctLearned{} with unpaired \ac{I2I}.
In the second step, we emulate the learned distortion by $\distFctLearned$ in the training dataset.
In the third step, we adapt the perception algorithm pre-trained with pristine data by fine-tuning with our adapted training dataset containing the learned distortion to improve the target domain performance.
It is noteworthy that our method does not require access to annotated training data in the target domain.

\vspace*{-.4em}
\subsection{Unpaired Learning of Unknown Image Distortions}
\label{sec:method_unpaired}
%
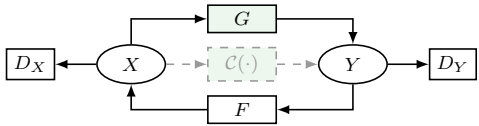
\begin{figure}
	\centering
	\input{images/cyclegan_blockDiag.tex}
	\vspace*{-1.4em}
	\caption{Unpaired learning of unknown image distortions from pristine source domain \undistDom{} to distorted target domain \distDom{}. Pristine-to-distortion mapping function \distFct{} (unattainable in practice) is approximated by forward generator $G$.}
	\vspace*{-1.9em}
	\label{fig:cyclegan_blockDiag}
\end{figure}
Our framework builds upon \discrThreeLay{}~\cite{Gruber2022}. We propose a \acf{SD} and an \acf{ESD} to study the assumption that shallow discriminators improve unpaired learning of unknown image distortions by better capturing local distortion characteristics.
We utilize {CycleGAN}~\cite{Zhu2017} for unpaired \ac{I2I}.
CycleGAN consists of two generators ($\generatorForward$, $\generatorBackward$) and two discriminators ($\discriminatorUndist$, $\discriminatorDist$).
As illustrated in \autoref{fig:cyclegan_blockDiag}, the forward generator $\generatorForward$ is trained to translate images from $\undistDom$ to $\distDom$, while the backward generator is trained for the inverse mapping $\generatorBackward : \distDom \rightarrow \undistDom$.
The forward generator $G$ represents the learned mapping \distFctLearned{}, i.e., the objective of our method.
The discriminator $\discriminatorUndist$ judges whether the given image is pristine or generated by $\generatorBackward$. 
Discriminator $\discriminatorDist$ judges whether an image is subject to the true or learned pristine-to-distorted mapping function.

The adversarial training consists of a least-squares \acs{GAN} loss $\mathcal{L}_\mathrm{GAN}$, a cycle-consistency loss $\mathcal{L}_\mathrm{c}$ and an identity mapping loss $\mathcal{L}_\mathrm{i}$.
The full objective can be formulated as:
\begin{equation}
	\begin{aligned}
		\mathcal{L}(G,F,D_X,D_Y) &= 
		\mathcal{L}_\mathrm{GAN} (G, D_Y) 
		+ \mathcal{L}_\mathrm{GAN} (F, D_X) \\
		&+ \lambda_\mathrm{c} \, \mathcal{L}_\mathrm{c} (G, F)
		+ \lambda_\mathrm{i} \, \mathcal{L}_\mathrm{i} (G, F) 
	\end{aligned}
\end{equation}
$\lambda_\mathrm{c}$ and $\lambda_\mathrm{i}$ denote the weighing of the respective losses.
The loss $\mathcal{L}_\mathrm{c}$ minimizes the error between input images $\myvec{x} \in \undistDom$ and their reconstruction after being passed through both generators $\hat{\myvec{x}} = \generatorBackward(\generatorForward(\myvec{x}))$ and vice versa.
The loss $\mathcal{L}_\mathrm{i}$ encourages color preservation in the generators by aiming for an identity mapping if an image already subject to the true distortion is fed into $G$ or a pristine image is fed into $F$. 
Further details on the losses and training procedure may be found in~\cite{Zhu2017}. We adopt the generators consisting of nine residual blocks from~\cite{Zhu2017}.

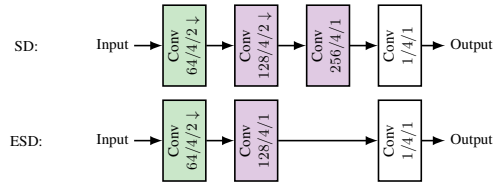
\begin{figure}
	\centering
	\input{images/discr-architecture.tex}
	\vspace*{-1.4em}
	\caption{Network architecture of discriminators. Conv $c/k/s \downarrow$ denotes a convolutional layer with $c$ output channels, a kernel size of $k$ and a stride of $s$. All convolutional layers marked in purple are followed by BatchNorm and Leaky ReLu. All layers marked in green are followed by Leaky ReLu.} 
	\vspace*{-1.9em}
	\label{fig:discr-architecture}
\end{figure}

For the discriminators, we introduce \ac{SD} and \ac{ESD} illustrated in \autoref{fig:discr-architecture}, which are $34 \times 34$ and $16 \times 16$ PatchGANs~\cite{Isola2017}, respectively.
PatchGANs convolutionally classify overlapping image patches and average the results to a single output.
The patch size equals the receptive field size of the last layer of the discriminator. 
The receptive field size is the area of the input image that is connected to the regarded output unit~\cite{Goodfellow2016_cnn}.
With each convolutional layer, the receptive field size of a network architecture increases.
The size of the discriminators' receptive field has a direct impact on its ability to capture local and global features.
%

Previous methods like \discrThreeLay{} exhibit limitations in learning local distortion characteristics.
We expect discriminators with smaller receptive fields to be better suited to capture local distortion characteristics.
We compute the receptive field size recursively for the $m$-th network layer as~\cite{Peng2022}:
\begin{equation}
	\label{eq:receptive-field}
	r_m = r_{m-1} + \left( k_m - 1 \right) \prod\nolimits_{i=1}^{m-1} s_i \mathrm{, where}\ r_0 = 1,
\end{equation}
$k_m$ and $s_m$ denote kernel size and stride of the convolution in the $m$-layer. It may be noted that this computation does not cover dilated convolutions.
A positive side effect of employing shallow discriminators is the low network complexity.
In \autoref{tab:discriminators}, we report receptive field size and complexity in number of parameters of our proposed discriminators alongside state of the art methods.
\begin{table}
	\vspace*{1.8em}
	\centering
	\caption{Discriminator receptive field size and complexity.}
	\label{tab:discriminators}
	\begin{tabular}{ccc}
		\toprule
		Discriminator &
		\parbox{3cm}{\centering Receptive Field Size} & \parbox{3cm}{\centering Complexity (\#Parameters)} \\
		\midrule
		\discrThreeLay{}~\cite{Gruber2022} & $70 \times 70$ & \num{2.765e6} \\
		\arruda{}~\cite{Arruda2022} & $16 \times 16$ & \num{0.136e6} \\
		\midrule
		SD (ours) & $34 \times 34$ & \num{0.663e6} \\
		ESD (ours) & $16 \times 16$ & \num{0.136e6} \\
		\bottomrule
		\vspace*{-2.5em}
	\end{tabular}
\end{table}

We aim at obtaining a framework that is capable of learning a wide range of distortions at arbitrary strength. 
Therefore, we employ the same parameter set for the learning of all considered distortions
for our proposed discriminator architectures.
This entails a batch size of \num{1} and a constant learning rate of \num{0.0002} during the first \num{100} epochs that linearly decays to zero over the following \num{100} epochs.
The training data is randomly cropped to $256 \times 256$ pixels and the losses are weighted by $\lambda_\mathrm{c} = 10$ and $\lambda_\mathrm{i} = 0.5$.

\vspace*{-.8em}
\subsection{Adaptation of Perception Algorithm}
\label{sec:method_adapt}
%
\begin{figure}
	\centering
	\input{images/framework_blockDiag.tex}
	\vspace*{-1.3em}
	\caption{Training procedure for pixel-level domain adaptation. True pristine-to-distorted mapping function $\distFct$ denotes mapping from pristine source domain $\undistDom$ to distorted target domain $\distDom$. $\distFctLearned$ denotes learned mapping.}
	\vspace*{-1.7em}
	\label{fig:framework_blockDiag}
\end{figure}
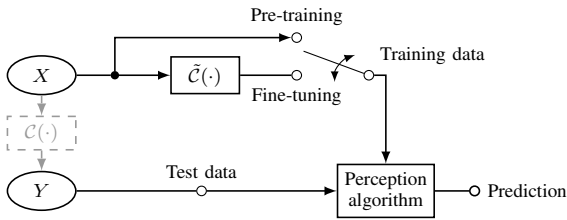
For each distorted domain, we establish a dedicated perception model. 
Specifically, we treat each distortion type and its individual distortion levels as separate domains $\distDom$ with their unique pristine-to-distorted mapping function $\distFct$.
For the dataset subject to \distFctLearned{}, the distortion is learned as described in \autoref{sec:method_unpaired} from validation and test dataset, and emulated in the training dataset as proposed in~\cite{Gruber2022}.
For our approach, we do not need annotated data from the target domain or additional datasets.
Finally, we adapt the perception algorithms to the desired image distortion by fine-tuning a model pre-trained on pristine data as illustrated in \autoref{fig:framework_blockDiag}.
True pristine-to-distortion mapping functions $\distFct$ shown in gray are employed only for evaluation, since they are unattainable in practice.

\section{Experimental Results}
\vspace*{-.4em}
\subsection{Experimental Setup}
\label{sec:experimental_setup}
\vspace*{-.4em}
We evaluate \ac{DA} for instance segmentation, because by encompassing detection and segmentation it is more challenging than other perception tasks.
We employ the Cityscapes dataset~\cite{Cordts2016} as pristine domain data due to its high quality.
For instance segmentation, we train Mask R-CNN~\cite{He2020} with a ResNet50~\cite{He2016} backbone with \texttt{Detectron2}~\cite{Wu2019}.
We adopt a \mbox{Mask R-CNN} model by~\cite{Wu2019}, which is pre-trained on COCO~\cite{Lin2014} and pristine Cityscapes.
During fine-tuning, the batch size is \num{4}, iterations are limited to \num{24000} and the learning rate of \num{0.01} is decreased to \num{0.001} after \num{18000} iterations.
All models are tested on the validation split degraded by the evaluated true distortion $\distFct$.
To compute the \ac{mAP} we follow~\cite{Cordts2016}.
We evaluate the following \ac{DA} scenarios:\\
\textbf{Baseline}: Model pre-trained on pristine data without adaptation to the unknown image distortion.\\
\textbf{Oracle}: Models fine-tuned with data subject to true pristine-to-distorted mapping function $\distFct$. This ideal scenario is unattainable in practice and serves as an upper bound.\\
\textbf{Proposed method and state of the art}: Models fine-tuned with data subject to learned pristine-to-distorted mapping function $\distFctLearned$. This entails \discrThreeLay{}~\cite{Gruber2022}, \arruda{}~\cite{Arruda2022}, and our \ac{SD} and \ac{ESD}.

The level of the evaluated distortions is controlled by $\blurQP$ of the Gaussian kernel for blur, $\noiseQP$ of the zero-mean Gaussian distribution for \ac{AWGN}, \ac{PSNR} for JPEG2000, \ac{CL} for \acs{JPEG}, and \ac{QP} for \ac{HEIF}.
In addition, we establish a simplified camera model to extend our method to learning of combined distortions leading to camera-specific image characteristics.
In the camera model, we apply blur, \ac{AWGN} and JPEG to simulate camera optics, sensor, and typical on-chip processing, respectively.
The parameters of our camera models are listed in \autoref{tab:simple_cam} alongside the mean \ac{PSNR} over the respective model dataset.
\begin{table}
	\centering
	\caption{Parameters for simplified camera models.}
	\label{tab:simple_cam}
	\begin{tabular}{ccccc}
		\toprule
		Model Identifier & \blurQP{} & \noiseQP{} & \jpegQP{} & Mean PSNR in dB \\
		\midrule
		A & \num{1} & \num{5}   & \num{34} & \num{35.50} \\
		B & \num{1} & \num{5}  & \num{26} & \num{34.86} \\
		C & \num{1} & \num{10}  & \num{34} & \num{33.57} \\
		D & \num{3} & \num{5}  & \num{34} & \num{30.69} \\
		E & \num{3} & \num{10}  & \num{34} & \num{29.91} \\
		F & \num{3} & \num{10}  & \num{18} & \num{29.82} \\
		\bottomrule
	\end{tabular}
	\vspace*{-2em}
\end{table}

\vspace*{-1.6em}
\subsection{Results for Individual Unknown Image Distortions}
\label{sec:results_individual}
%
\begin{figure*}[t]
	\centering
	\input{images/all_aps_tikzpicture.tex}
	\vspace*{-1.4em}
	\caption{Results of instance segmentation measured as \acf{mAP} over distortion level for various distortion types.}
	\vspace*{-1.9em}
	\label{fig:results_ap}
\end{figure*}
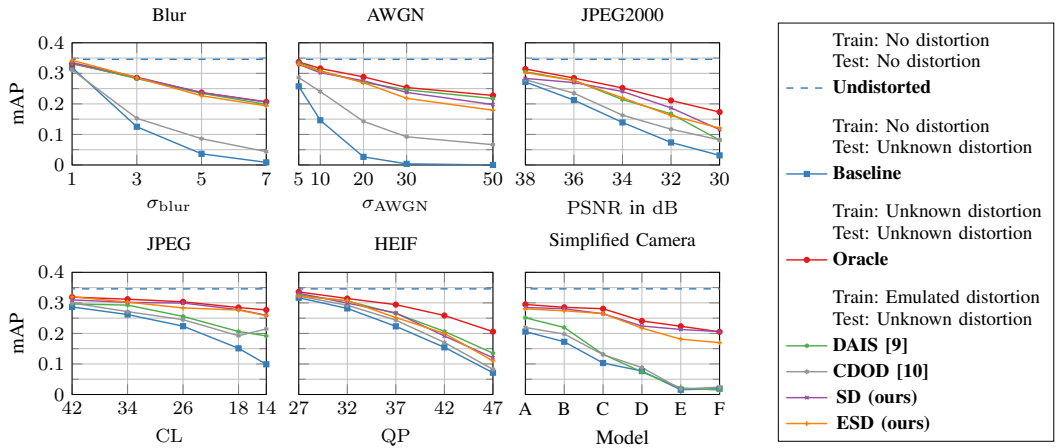
The results in terms of \ac{mAP} over the distortion level are shown in \autoref{fig:results_ap}. 
For all methods and distortions, the \ac{mAP} decreases with an increasing level of distortion and the oracle illustrates the ideal case.
For all scenarios, \arruda{}~\cite{Arruda2022} yields the lowest performance.
A possible explanation for this is that the resizing of the training images alters the distortion characteristics leading to limitations of this approach.

For blur and \ac{AWGN}, \discrThreeLay{}~\cite{Gruber2022} and our \discrTwoLay{} and \discrOneLay{} perform close to the oracle.
For \ac{AWGN} and our \discrTwoLay{} and \discrOneLay{}, we see a slight deterioration at distortion levels $\noiseQP \geq 30$.
The probable reason is the discriminator seeing less randomness per image patch with a narrower receptive field.
For JPEG2000 and JPEG, we see that with decreasing \ac{PSNR} and \ac{CL}, our \discrTwoLay{} and \discrOneLay{} clearly outperform \discrThreeLay{}~\cite{Gruber2022}.
These results support our assumption that smaller receptive fields improve learning of local distortion characteristics, since these become predominant for stronger coding.
For JPEG2000, these predominant artifacts are general blurring caused by the quantization of wavelet coefficients and ringing occurring at edges.
For JPEG, blocking caused by the fixed coding block size and color bleeding due to 	chroma subsampling are predominant.
For \ac{HEIF}, all discriminators achieve similar \ac{mAP} values.
While these results exhibit an improvement over the baseline, reaching the oracle continues to be a challenge.
This could be due to stronger content-dependency of the artifacts in \ac{HEIF} coding, which arises from the use of adaptive coding block sizes and more complex coding procedures in in-loop filters, prediction modes, and rate-distortion optimization.
The image quality of strongly compressed \ac{HEIF} images is substantially higher than that of JPEG or JPEG2000, resulting in fewer visually disturbing artifacts.
We also investigated a larger discriminator (receptive field size $142 \times 142$), which achieved results comparable to \discrThreeLay{}~\cite{Gruber2022} for blur, JPEG, and lower levels of \ac{AWGN}, JPEG2000, and HEIF distortion. However, due to inferior performance at stronger distortion levels  of \ac{AWGN}, JPEG2000, and HEIF it is not included in this work.

As shown in \autoref{tab:discriminators}, our approach greatly reduces the complexity compared to \discrThreeLay{}~\cite{Gruber2022}.
The number of parameters is reduced approximately by a factor of \num{4} for our \discrTwoLay{} and \num{20} for our \discrOneLay{}. 
At this decreased complexity, we outperform \discrThreeLay{} with our \ac{ESD} for JPEG by \num{0.07} and JPEG2000 by \num{0.04} in terms of \ac{mAP} and retain their performance for blur, \ac{AWGN} and \ac{HEIF}.
Our \ac{ESD} matches the complexity of \arruda{}~\cite{Arruda2022} in terms of number of parameters, while our SD involves approximately five times more parameters.
Our \discrOneLay{} clearly outperforms \arruda{} in terms of \ac{mAP} with maximum gains ranging from \numrange{0.03}{0.15}, while retaining the same complexity.

We can conclude that our proposed shallow discriminators \ac{SD} and \ac{ESD} achieve state of the art performance in pixel-level \ac{DA} for instance segmentation.
We clearly outperform \arruda{}~\cite{Arruda2022} at the same complexity.
In the presence of blur and \ac{HEIF} distortion, we achieve the same performance as \discrThreeLay{}~\cite{Gruber2022} at a reduced complexity.
For JPEG2000 coding, slight, and for \acs{JPEG}, vast improvements over \discrThreeLay{} are observed.
Our approach improves learning of local distortion characteristics, which are predominant in these cases.
For strong \ac{AWGN}, we observe slightly lower \acp{mAP} than \discrThreeLay{} with shallow discriminators.
However, in the light of the gains obtained for \acs{JPEG} and JPEG2000 over \discrThreeLay{} and the substantial gains over \arruda{}, we consider this to be a manageable trade-off that does not compromise the overall effectiveness of our approach.
\discrTwoLay{} and \discrOneLay{} showed similar performance, with slightly better results for the \discrTwoLay{}.

\vspace*{-.6em}
\subsection{Results for Camera-Specific Image Characteristics}
In real-world scenarios, distortions typically occur in combination rather than in isolation.
Therefore, we investigate the simplified camera model described in \autoref{sec:experimental_setup} with model parameters given in \autoref{tab:simple_cam}.
The results in terms of \ac{mAP} are given in \autoref{fig:results_ap}.
For all camera models, our \discrTwoLay{} and \discrOneLay{} reach \acp{mAP} close to the oracle and state of the art methods are clearly inferior. 
With our \discrOneLay{} we achieve maximum \ac{mAP} gains of \num{0.16} over both \discrThreeLay{}~\cite{Gruber2022} and \arruda{}~\cite{Arruda2022}.
These findings reinforce the results presented and discussed in \autoref{sec:results_individual} and support our proposal to use shallow discriminators.

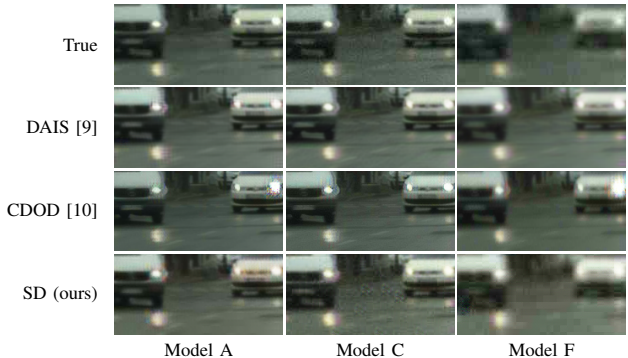
\begin{figure}
	\centering
	\input{images/visuals_simple_cam_transposed.tex}
	\vspace*{-1.4em}
	\caption{Visual examples of true and learned camera-specific image characteristics in Cityscapes image \texttt{dusseldorf\_000028\_000019} for simplified camera models A, C, and F. \textit{(Best to be viewed enlarged on a monitor.)}}
	\vspace*{-1.7em}
	\label{fig:visuals_simple_cam}
\end{figure}
In \autoref{fig:visuals_simple_cam}, visual examples for camera models A, C, and F are depicted. 
Apparently, the distortions learned with the state of the art methods \discrThreeLay{}~\cite{Gruber2022} and \arruda{}~\cite{Arruda2022} are significantly weaker than the true distortion. 
Our \discrTwoLay{} enables capturing these distortions.
In particular, blocking and quantization artifacts not captured before are clearly visible for all camera models, most prominently in Model F.
It seems that without resizing of training images as proposed in \arruda{} and with a shallower discriminators camera-specific image characteristics can be captured more reliably.

\vspace*{-.8em}
\subsection{Ablation Study}
\label{sec:results_cut}
We assess the suitability of our shallow discriminators \ac{SD}-A and \ac{ESD}-A compared to a \ac{BD-A} with another \ac{I2I} approach. 
Specifically, we replace concepts from CycleGAN~\cite{Zhu2017} by CUT~\cite{Park2020} within our proposed method for unpaired learning of unknown image distortion.
The \ac{BD-A} with a $70 \times 70$ receptive field is comparable to \discrThreeLay{}~\cite{Gruber2022}.
The same network architectures and hyperparameters as in our proposed method (\autoref{sec:method_unpaired}) are used, with CUT losses weighed by $\lambda_\mathrm{GAN}=1.0$ and $\lambda_\mathrm{NCE} = 1.0$.
We focus on our simplified camera model, as the occurring camera-specific image characteristics are more complex and practically relevant than individual distortions. 

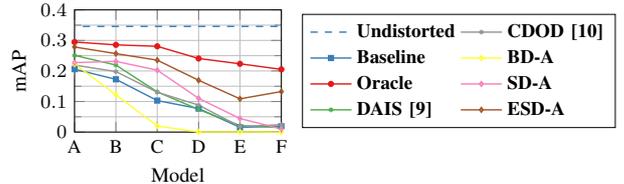
\begin{figure}
	\centering
	\input{images/ablation_aps_tikzpicture.tex}
	\vspace*{-1.4em}
	\caption{Ablation results of instance segmentation for simplified camera model.}
	\vspace*{-1.7em}
	\label{fig:ablation_results_ap}
\end{figure}
As shown in \autoref{fig:ablation_results_ap}, CUT is less suitable for learning camera-specific image characteristics.
However, the ablation discriminators \ac{SD}-A and \ac{ESD}-A still outperform both the \ac{BD-A} and state of the art methods \discrThreeLay{}~\cite{Gruber2022} and \arruda{}~\cite{Arruda2022}.
This demonstrates the generalizability of our proposed shallow discriminators and supports our hypothesis that discriminators with smaller receptive field sizes are more suitable for capturing local distortion characteristics.

\section{Conclusion}
\label{sec:conclusion}
In this paper, we proposed two shallow discriminators for improved learning of unknown image distortions leading to camera-specific image characteristics.
We demonstrated that our shallow discriminators \ac{SD} and \ac{ESD} are better suited to reproduce local distortion characteristics due to their smaller receptive fields and their low computational complexity.
Our findings are quantified for \ac{DA} in instance segmentation and verified with visual examples.
In an ablation study, we showed the generalizability of using shallow discriminators for a different \ac{I2I} method within our unpaired learning approach.
In future work, \ac{HEIF} coding should be investigated in more detail for further improvement. 
Furthermore, our method could be extended to incorporate more sophisticated lens distortion models through spatial-dependency, and more accurate sensor noise by addressing amplitude-dependency. 

\section*{Acknowledgment}
We gratefully acknowledge support by the Bavarian Ministry for Economic Affairs, Regional Development and Energy (StMWi) under Grant No. DIK0179/02

%
%
\bibliographystyle{IEEEtran}
\bibliography{IEEEabrv,literature.bib}
%
%
%
%
%
\end{document}

%% file: images/unknown_distortion.tex
\begin{tikzpicture}[node distance=0.4cm and 0.4cm, scale=0.84, transform shape,
	distbox/.style={draw=none, rectangle, rounded corners, align=center, fill=myblue!10, minimum width=2.9cm},
	stagebox/.style={draw, rectangle, fill=myyellow!10, thick, align=center, minimum height=1.1cm, minimum width=3.6cm},
	arrow/.style={->, thick},
	dashedarrow/.style={->, thick, dashed},
	dashedbox/.style={draw=myblue, dashed, thick, rectangle}, 
	]
	
	\node (scene) [align=center] {Real World\\Scene};
	\node (optics) [stagebox, below=of scene] {\textbf{Optics}\\Lens\\Aperture\\};
	\node (sensor) [stagebox, below=of optics] {\textbf{Sensor}\\Color Filter\\Spatial Sampling\\A/D Conversion};
	\node (processing) [stagebox, below=of sensor] {\textbf{On-Chip Image Processing}\\Demosaicing\\Image Enhancement\\Compression};
	\node (image) [below=of processing] {Bitstream};
	
	\draw[arrow] (scene) -- (optics);
	\draw[arrow] (optics) -- (sensor);
	\draw[arrow] (sensor) -- (processing);
	\draw[arrow] (processing) -- (image);
	
	\node (optics_dist) [distbox, right=of optics] {
		- Blur\\
		- Optical Distortion\\
		- Chromatic Aberration
	};
	
	\node (sensor_dist) [distbox, right=of sensor] {
		- Registration Noise\\			
		- Quantization Noise\\
		- Sampling Artifacts
	};
	
	\node (processing_dist) [distbox, right=of processing] {
		- Processing Artifacts\\
		- Coding Artifacts\\
	};
	
	\node[left=.1cm of optics.west] (imOptics) {\includegraphics[trim=85 80 10 30,clip,width=1.8cm]{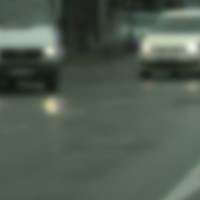}};
	\node[left=.1cm of sensor.west] (imSensor) {\includegraphics[trim=85 80 10 30,clip,width=1.8cm]{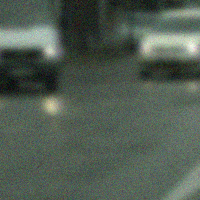}};
	\node[left=.1cm of processing.west] (imProcessing) {\includegraphics[trim=85 80 10 30,clip,width=1.8cm]{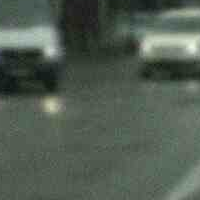}};
	
	\node[above=.18cm of imOptics.north] (imInput) {\includegraphics[trim=85 80 10 30,clip,width=1.8cm]{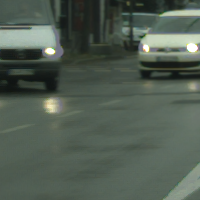}};
	
	\draw[dashedbox] ($(optics_dist.north west) + (-0.15, 0.15)$) rectangle ($(processing_dist.south east) + (0.15, -0.15)$);
	\node[text width=2.8cm, align=center] at ($(processing_dist.south) + (0,-.7)$) { {\textbf{\textcolor{myblue}{Unknown Individual Distortions}}}};
	\draw[thick,myblue,decorate,decoration={brace,amplitude=12pt}, transform canvas={xshift=6pt}] (optics_dist.north east) -- (processing_dist.south east);
	\node[right=of processing_dist.south east, rotate=90, yshift=-.6cm, xshift=-.15cm] {\textbf{\textcolor{myblue}{{Camera-Specific Image Characteristics}}}};

\end{tikzpicture}

%% file: images/cyclegan_blockDiag.tex
\begin{tikzpicture}[framed, background rectangle/.style={thick,draw=none}, scale=0.9, transform shape]
	\definecolor{mygreen}{HTML}{4daf4a}
	\node[source, minimum width=1cm, black, semithick] (undistDom) {\undistDom{}};
	\node[system, right=.6cm of undistDom, minimum width=1cm, dashed, mygray, semithick, fill=mygreen!10] (distFct) {$\distFct(\cdot)$};
	\node[source, minimum width=1cm, black, right=.6cm of distFct, semithick] (distDom) {\distDom{}};
	
	\draw[->, dashed, mygray, semithick] (undistDom) -- (distFct);
	\draw[->, dashed, mygray, semithick] (distFct) -- (distDom);
	
	\node[system, right=.6cm of distDom, semithick] (discrDist) {$D_Y$};
	\node[system, left=.6cm of undistDom, semithick] (discrUndist) {$D_X$};
	\draw[->, semithick] (distDom) -- (discrDist);
	\draw[->, semithick] (undistDom) -- (discrUndist);
	
	\node[system, above=.2cm of distFct, semithick, minimum width=1cm, fill=mygreen!10] (genForw) {$G$};
	\node[system, below=.2cm of distFct, semithick, minimum width=1cm] (genBack) {$F$};
	\draw[->, semithick] (undistDom) |- (genForw);
	\draw[->, semithick] (genForw) -| (distDom);
	\draw[->, semithick] (distDom) |- (genBack);
	\draw[->, semithick] (genBack) -| (undistDom);
\end{tikzpicture}

%% file: images/discr-architecture.tex
\begin{tikzpicture}[node distance=1.5cm, every node/.style={scale=0.7}, framed, background rectangle/.style={thick,draw=none}, scale=0.9, transform shape]
	\tikzstyle{box} = [draw]
	\definecolor{colorConv1}{HTML}{4daf4a} 
	\definecolor{colorConv2}{HTML}{984ea3} 
	\definecolor{colorPool}{HTML}{999999} 
	
	\node[align=center,semithick] (discr2-input) at (0, 0) {Input};
	\node[box,semithick,align=center, minimum width=1.7cm, right of=discr2-input, fill=colorConv1!30, rotate=90](discr2-lay1){Conv\\$64/4/2 \downarrow$};
	\node[box,semithick,align=center, minimum width=1.7cm, right of=discr2-lay1, fill=colorConv2!30, rotate=90](discr2-lay2){Conv\\$128/4/2 \downarrow$};
	\node[box,semithick,align=center, minimum width=1.7cm, right of=discr2-lay2, fill=colorConv2!30, rotate=90](discr2-lay3){Conv\\$256/4/1$};
	\node[box,semithick,align=center, minimum width=1.7cm, right of=discr2-lay3, fill=white, rotate=90](discr2-lay6){Conv\\$1/4/1$};
	\node[align=right,semithick, right of=discr2-lay6](discr2-output) {Output};
	\draw[->,semithick] (discr2-input) --(discr2-lay1);
	\draw[->,semithick] (discr2-lay1) -- (discr2-lay2);
	\draw[->,semithick] (discr2-lay2) -- (discr2-lay3);
	\draw[->,semithick] (discr2-lay3) -- (discr2-lay6);
	\draw[->,semithick] (discr2-lay6) -- (discr2-output);
	
	\node[align=center,semithick, below of=discr2-input, yshift=-.5cm] (discr1-input) {Input};
	\node[box,semithick,align=center, minimum width=1.7cm, right of=discr1-input, fill=colorConv1!30, rotate=90](discr1-lay1){Conv\\$64/4/2 \downarrow$};
	\node[box,semithick,align=center, minimum width=1.7cm, right of=discr1-lay1, fill=colorConv2!30, rotate=90](discr1-lay2){Conv\\$128/4/1$};
	\node[box,semithick,align=center, minimum width=1.7cm, below of=discr2-lay6, yshift=-.5cm, fill=white, rotate=90](discr1-lay6){Conv\\$1/4/1$};
	\node[align=right,semithick, right of=discr1-lay6](discr1-output) {Output};
	\draw[->,semithick] (discr1-input) --(discr1-lay1);
	\draw[->,semithick] (discr1-lay1) -- (discr1-lay2);
	\draw[->,semithick] (discr1-lay2) -- (discr1-lay6);
	\draw[->,semithick] (discr1-lay6) -- (discr1-output);
	
	\node[align=center,semithick, left of=discr1-input, xshift=-0.3cm] (discr1-name) {ESD:};
	\node[align=center,semithick, left of=discr2-input, xshift=-0.3cm] (discr2-name) {SD:};	
\end{tikzpicture}

%% file: images/framework_blockDiag.tex
\begin{tikzpicture}[framed, background rectangle/.style={thick,draw=none}, scale=0.9, transform shape]
	\node[source, minimum width=1cm, black, semithick] (undistDom) {\undistDom{}};
	\node[system, below=.3cm of undistDom, minimum width=1cm, dashed, mygray, semithick] (distFct) {$\distFct(\cdot)$};
	\node[source, minimum width=1cm, black, below=.3cm of distFct, semithick] (distDom) {\distDom{}};
	
	\draw[->, dashed, mygray, semithick] (undistDom) -- (distFct);
	\draw[->, dashed, mygray, semithick] (distFct) -- (distDom);
	
	\node[syssplit, right=.5cm of undistDom, semithick] (split) {};
	\draw[-, semithick] (undistDom) -- (split);

	\node[system,right=.75cm of split,align=center, minimum width=1cm, semithick](distFctLearned){$\distFctLearned(\cdot)$};
	\draw[->, semithick] (split) -- (distFctLearned);
	\node[syscon,right=.75cm of distFctLearned, label={[align=center]below:{Fine-tuning}}](distFctLearnedCon){};
	\draw[-, semithick] (distFctLearned) -- (distFctLearnedCon);

	\node[syscon,above=.4cm of distFctLearnedCon, label={[align=center]above:{Pre-training}}](noFTCon){};
	\draw[->, semithick] (split) |- (noFTCon);
	
	\node[syscon,right=.9cm of distFctLearnedCon, label={[align=left]above right:{Training data}}](casesEnd){};

	\node[syscon,right=1.75cm of distDom, label={[align=center]above:{Test data}}](testData){};
	\node[system,right=1.9cm of testData,align=center, minimum width=1cm, semithick](instSeg){Perception\\algorithm};
	\draw[->, semithick] (casesEnd) -| (instSeg) {};
	\draw[-, semithick] (distDom) -- (testData) {};
	\draw[->, semithick] (testData) -- (instSeg) {};
	
	\node[syscon, right=.5cm of instSeg, label={[align=center] right:{Prediction}}, semithick] (output) {};
	\draw [semithick] (instSeg) -- (output);
	
	\coordinate[above left= .3 and .9 of casesEnd] (switch);
	\draw (casesEnd) -- (switch);
	\draw[<->] (casesEnd) ++(-0.5, -0.1) to[out=90, in=-160] ++(.3, .4);

\end{tikzpicture}

%% file: images/all_aps_tikzpicture.tex
\definecolor{col1}{RGB}{55,126,184}
\definecolor{col2}{RGB}{228,26,28}
\definecolor{col3}{RGB}{77,175,74}
\definecolor{col4}{RGB}{152,78,163}
\definecolor{col5}{RGB}{255,127,0}
\definecolor{col6}{RGB}{150,150,150}
\definecolor{col7}{RGB}{255,255,51}
\definecolor{col8}{RGB}{247,129,191}
\definecolor{col9}{RGB}{166,86,40} 
\begin{tikzpicture}[framed, background rectangle/.style={thick,draw=none}, scale=0.9, transform shape]
 	\begin{groupplot}[
 		group style={
 			group size=3 by 2,
 			group name=allAps,
 			x descriptions at=edge bottom,
 			y descriptions at=edge left,
 			horizontal sep=1.5em,
 			vertical sep=5em
			},
		width=.24\textwidth,
		ymin=0,
		ymax=0.40,
		ytick={0,0.1,...,0.4},
		ylabel={mAP},
		y=4.5cm,
		xtick=data,
		enlarge x limits=false,
		grid=both,
		minor x tick num=0,
		minor y tick num=1,
 		/tikz/font=\footnotesize,
 		legend cell align=left,
 		]
 		\input{images/blur_gauss_ap_group.tex}
		\input{images/noise_gauss_ap_group.tex}
		\input{images/j2k_ap_group.tex}
		\input{images/jpeg_ap_group.tex}
		\input{images/heif_ap_group.tex}
		\input{images/simple_cam_ap_group.tex}
	\end{groupplot}
\end{tikzpicture}

%% file: images/visuals_simple_cam_transposed.tex
\newcommand{\mywidth}{.12\textwidth}
\newcommand{\myfontsize}{\scriptsize}
\newcommand{\impath}[5]{../crops_dusseldorf/#1_qp_#2_#3_#4_#5_max.png}
%
%
\begin{tikzpicture}[scale=0.85, framed, background rectangle/.style={thick,draw=none}]
	\newcommand{\imname}{dusseldorf_000028_000019}
	\newcommand{\cropnum}{1}
	\input{images/visuals_template_simple_cam_transposed.tex}
\end{tikzpicture}

%% file: images/ablation_aps_tikzpicture.tex
\definecolor{col1}{RGB}{55,126,184}
\definecolor{col2}{RGB}{228,26,28}
\definecolor{col3}{RGB}{77,175,74}
\definecolor{col4}{RGB}{152,78,163}
\definecolor{col5}{RGB}{255,127,0}
\definecolor{col6}{RGB}{150,150,150}
\definecolor{col7}{RGB}{255,255,51}
\definecolor{col8}{RGB}{247,129,191}
\definecolor{col9}{RGB}{166,86,40} 
\begin{tikzpicture}[framed, background rectangle/.style={thick,draw=none}, scale=0.9, transform shape]
 	\begin{groupplot}[
 		group style={
 			group size=1 by 1,
 			group name=allAps,
 			x descriptions at=edge bottom,
 			y descriptions at=edge left,
 			horizontal sep=0em,
 			vertical sep=0em
			},
		width=.25\textwidth,
		ymin=0,
		ymax=0.40,
		ytick={0,0.1,...,0.4},
		ylabel={mAP},
		y=4.5cm,
		xtick=data,
		enlarge x limits=false,
		grid=both,
		minor x tick num=0,
		minor y tick num=1,
 		/tikz/font=\footnotesize,
 		legend cell align=left,
 		]
		\input{images/ablation_simple_cam_ap_group.tex}
	\end{groupplot}
\end{tikzpicture}